\documentstyle[aps,12pt,amssymb]{revtex}

\begin{document}
\baselineskip=0.75cm
\newcommand{\ini}{\begin{equation}}
\newcommand{\fin}{\end{equation}}
\newcommand{\inir}{\begin{eqnarray}}
\newcommand{\finr}{\end{eqnarray}}
\newcommand{\inif}{\begin{figure}}
\newcommand{\finf}{\end{figure}}
\newcommand{\bc}{\begin{center}}
\newcommand{\ec}{\end{center}}
\def\ol{\overline}
\def\pa{\partial}
\def\ra{\rightarrow}
\def\ts{\times}
\def\df{\dotfill}
\def\bs{\backslash}
\def\dg{\dagger}

$~$

\hfill DSF-17/2001

\vspace{1 cm}

\centerline{{\bf MODEL FOR FERMION MASS MATRICES AND}} 
\centerline{{\bf THE ORIGIN OF QUARK-LEPTON SYMMETRY}}

\vspace{1 cm}

\centerline{\large{D. Falcone}}

\vspace{1 cm}

\centerline{Dipartimento di Scienze Fisiche, Universit\`a di Napoli,}
\centerline{Complesso di Monte S. Angelo, Via Cintia, Napoli, Italy}
\centerline{e-mail: falcone@na.infn.it}

\vspace{1 cm}

\begin{abstract}

\noindent
Several phenomenological features of fermion masses and mixings can
be accounted for by a simple model for fermion mass matrices, which suggests
an underlying U(2) horizontal symmetry. In this context, it is also proposed
how an approximate quark-lepton symmetry can be achieved without unified
gauge theories.

\end{abstract}

\newpage

Understanding the pattern of fermion masses and mixings is a key subject
in modern particle physics. Many approaches have been followed \cite{many}.
For example, discrete and continuous as well as abelian and nonabelian
horizontal symmetries have been often used. Moreover, unified
gauge theories, which are based on vertical symmetries,
generally predict relations between quark and lepton masses \cite{gj}.
Horizontal symmetries relate particles of different generations, while
vertical symmetries relate particles of the same generation. Even the
$SU(3) \ts SU(2) \ts U(1)$ symmetries of the standard model or the
$SU(3) \ts SU(2) \ts SU(2) \ts U(1)$ symmetries of the left-right model
are vertical. In this paper we show that several phenomenological
issues, among which quark-lepton symmetry, seesaw mechanism and large
neutrino mixing, lead to a simple model for quark and lepton mass
matrices, suggesting an underlying broken $U(2)$ horizontal symmetry 
\cite{u2}. Speculations on the origin of quark-lepton symmetry,
not relying on unified models, and hence not leading to proton 
instability, are also presented. 

It is well known that quark masses and mixings exhibit a hierarchical
pattern,
\ini
\frac{m_u}{m_c} \sim \frac{m_c}{m_t} \sim \lambda^4,
\fin
\ini
\frac{m_d}{m_s} \sim \frac{m_s}{m_b} \sim \lambda^2,
\fin
\ini
V_{us} \sim \lambda,~V_{cb} \sim \lambda ^2,~V_{ub} \sim \lambda ^4,
\fin
where $\lambda =0.22$ is the sine of the Cabibbo angle, and from Eqn.(3) we
see that quark mixings are small. Moreover, the hierarchy of
charged lepton masses is similar to that of down quark masses,
\ini
\frac{m_e}{m_{\mu}} \sim \frac{m_{\mu}}{m_{\tau}} \sim \lambda^2.
\fin
For the masses of the third generation we have
\ini
m_t \gg m_b \sim m_{\tau},
\fin
and the top quark mass is nearly equal to the vacuum expectation value (VEV)
of the standard model Higgs boson.

Quark masses and mixings arise when quark mass matrices are diagonalized
by means of biunitary transformations $V_{uL}^{\dg}M_u V_{uR} =D_u$,
$V_{dL}^{\dg}M_d V_{dR} =D_d$. The Cabibbo-Kobayashi-Maskawa (CKM) quark mixing
matrix \cite{ckm} is given by the formula $V_{CKM}=V_{uL}^{\dg} V_{dL}$. 
In recent papers \cite{cf} a parallel structure for up and down quark
mass matrices has been obtained,
\ini
M_u \simeq \left( \begin{array}{ccc}
0 & \sqrt{m_u m_c} & 0 \\ \sqrt{m_u m_c} & m_c & \sqrt{m_u m_t} \\
0 & \sqrt{m_u m_t} & m_t
\end{array} \right),
\fin
\ini
M_d \simeq \left( \begin{array}{ccc}
0 & \sqrt{m_d m_s} & 0 \\ \sqrt{m_d m_s} & m_s & \sqrt{m_d m_b} \\
0 & \sqrt{m_d m_b} & m_b
\end{array} \right).
\fin
By using Eqns.(1),(2) we get these mass matrices in terms of powers
of $\lambda$ and an overall mass scale,
\ini
M_u \sim \left( \begin{array}{ccc}
0 & \lambda^6 & 0 \\ \lambda^6 & \lambda^4 & \lambda^4 \\ 0 & \lambda^4 & 1
\end{array} \right)m_t,
\fin
\ini
M_d \sim \left( \begin{array}{ccc}
0 & \lambda^3 & 0 \\ \lambda^3 & \lambda^2 & \lambda^2 \\ 0 & \lambda^2 & 1
\end{array} \right)m_b.
\fin
This structure for $M_u$ and $M_d$ strongly suggests the presence of a
broken $U(2)$ horizontal symmetry \cite{bar1}, if we consider the zeros
as approximate. In particular, for symmetric forms,
the similarity $M_{22} \sim M_{23}$ favours the $U(2)$ symmetry,
rather than the $U(1)$ symmetry, where we have $M_{22} \sim (M_{23})^2$
(see for example Ref.\cite{ber}). Mixings come out mainly from the down sector,
where the hierarchy is less pronounced.

Now we have to consider lepton masses and mixings. In unified theories, such
as the $SO(10)$ model, the charged lepton mass matrix $M_e$ is related to
$M_d$ and the Dirac neutrino mass matrix $M_{\nu}$ is related to $M_u$,
\ini
M_e \sim \frac{m_{\tau}}{m_b}~M_d,
\fin
\ini
M_{\nu} \sim \frac{m_{\tau}}{m_b}~M_u,
\fin
where the factor $m_{\tau}/m_b$ takes into account approximate running of
quark masses with respect to lepton masses, and $m_b \simeq m_{\tau}$ at the
unification scale in the supersymmetric case \cite{ara}.
This is called quark-lepton symmetry. Moreover, the seesaw mechanism
\cite{seesaw} holds, according to the formula
\ini
M_L \simeq M_{\nu} M_R^{-1} M_{\nu},
\fin
where $M_R$ is the Majorana mass matrix of heavy right-handed neutrinos
and $M_L$ is the Majorana mass matrix of light left-handed neutrinos.
The seesaw mechanism can explain the smallness of the effective neutrino mass,
since the mass of the right-handed neutrino is generated at the unification
scale $M_U \sim 10^{16}$ GeV in the supersymmetric case and the Dirac mass
at the weak scale $M_W \sim 10^{2}$ GeV.
Lepton masses and mixings arise when $M_L$ and $M_e$ are diagonalized
through the transformations $V_L^{\dg}M_L V_L^* =D_L$,
$V_{eL}^{\dg}M_e V_{eR}=D_e$. The Maki-Nakagawa-Sakata (MNS) lepton mixing
matrix \cite{mns} is given by $V_{MNS}=V_L^{\dg} V_{eL}$.
In Eqns.(10),(11) we may neglect the running of quark mixings, which is not relevant
for our analysis. In fact, also the running of quark masses is unimportant.
However, we keep the factor $m_{\tau}/m_b$ to remember the presence of the
running effect, which is responsible for the mass difference between $m_b$
and $m_{\tau}$ at the low scale. Quark-lepton symmetry is due to the
$SU(4) \ts SU(2) \ts SU(2)$ subgroup of $SO(10)$ and in particular to the
$SU(4)$ component, which includes lepton number as fourth color \cite{ps}.
In the $SU(5)$ model, only relation (10) is achieved.

It can be shown \cite{fal} that the large mixing angle solutions to
the solar neutrino problem \cite{bks}, which are favoured by recent data fits
\cite{fuba}, along with the large mixing of atmospheric neutrinos,
give the remarkable result of a nearly democratic
\ini
M_L^{-1} \sim \left( \begin{array}{ccc}
1 & 1 & 1 \\ 1 & 1 & 1 \\ 1 & 1 & 1
\end{array} \right) \frac{1}{m_1},
\fin
where $m_1$ is the mass of the lightest left-handed neutrino in the normal
hierarchy case \cite{bgg}.
Then, by using Eqns.(8),(11),(12) and (13) we obtain
\ini
M_R \sim \left( \frac{m_{\tau}}{m_b} \right)^2
\left( \begin{array}{ccc}
0 & 0 & \lambda^6 \\ 0 & \lambda^8 & \lambda^4 \\
\lambda^6 & \lambda^4 & 1
\end{array} \right) \frac{m_t^2}{m_1},
\fin
where again the zeros are approximate.
The overall scale of $M_R$ is consistent with or higher than the unification
scale. Consistence is achieved for $m_1 \sim 10^{-4}$ eV.

Therefore, all fermion mass matrices can now be written in a form suggested by
the breaking of the $U(2)$ horizontal symmetry \cite{bar2,bar3,car}, and filling
the zero elements by higher powers of $\lambda$ we get
\ini
M_u \sim \left( \begin{array}{ccc}
(\lambda^6)^2 & \lambda^6 & \lambda^4 \lambda^6 \\
\lambda^6 & \lambda^4 & \lambda^4 \\
\lambda^4 \lambda^6 & \lambda^4 & 1
\end{array} \right)m_t \sim M_{\nu},
\fin
\ini
M_d \sim \left( \begin{array}{ccc}
(\lambda^3)^2 & \lambda^3 & \lambda^2  \lambda^3 \\
\lambda^3 & \lambda^2 & \lambda^2 \\
\lambda^2  \lambda^3 & \lambda^2 & 1
\end{array} \right)m_b \sim M_e,
\fin
\ini
M_R \sim \left( \begin{array}{ccc}
(\lambda^6)^2 & \lambda^4  \lambda^6 & \lambda^6\\
\lambda^4  \lambda^6 & (\lambda^4)^2 & \lambda^4 \\
\lambda^6 & \lambda^4 & 1
\end{array} \right)M_3,~~~~~~
\fin
with $M_3$ the mass of the heaviest right-handed neutrino. Coefficients are
greater than $\lambda$ and less than $1/\lambda$.
We stress that the heavy neutrino mass matrix is found, and not built,
consistent with the broken $U(2)$ symmetry, as a consequence of the democratic
form (13). The $U(2)$ breaking parameters are given by
\ini
\epsilon_u \simeq  {\epsilon_d}^2 \simeq \lambda^4,
\fin
\ini
\epsilon'_u \simeq {\epsilon'_d}^2 \simeq \lambda^6,
\fin
so that mass matrices can be written as
\ini
M_u \sim \left( \begin{array}{ccc}
{\epsilon'_u}^2 & \epsilon'_u & \epsilon_u  \epsilon'_u \\
\epsilon'_u & \epsilon_u & \epsilon_u \\
\epsilon_u  \epsilon'_u & \epsilon_u & 1
\end{array} \right)m_t \sim M_{\nu},
\fin
\ini
M_d \sim \left( \begin{array}{ccc}
{\epsilon'_d}^2 & \epsilon'_d & \epsilon_d  \epsilon'_d \\
\epsilon'_d & \epsilon_d & \epsilon_d \\
\epsilon_d  \epsilon'_d & \epsilon_d & 1
\end{array} \right)m_b \sim M_e,
\fin
\ini
M_R \sim \left( \begin{array}{ccc}
{\epsilon'_u}^2 & \epsilon_u  \epsilon'_u & \epsilon'_u \\
\epsilon_u  \epsilon'_u & {\epsilon_u}^2 & \epsilon_u \\
\epsilon'_u & \epsilon_u & 1
\end{array} \right)M_3.~~~~~~
\fin
The parameters $\epsilon$ are involved in the breaking of $U(2)$ to $U(1)$, and
the parameters $\epsilon'$ in the further breaking of this $U(1)$.
According to the $U(2)$ horizontal symmetry, the three generations $\psi_i$
transform as a doublet $\psi_a$ plus a singlet $\psi_3$. This is attractive
for supersymmetry, because leads to the degeneracy between first and second
generation, which is needed to suppress the flavor changing neutral currents
in the squark sector \cite{u2}.
Thus we discuss the origin of matrices in Eqns.(15),(16) and (17)
within the supersymmetric standard model with right-handed neutrinos.
We do not assume the existence of a unified model. In fact, in unified theories
with $U(2)$ as horizontal symmetry, the mechanisms
for the enhancement of the up quark mass hierarchy with respect to the down
quark mass hierarchy are quite involved \cite{bar1,car}. Moreover, it is not
clear if a simple Higgs structure is able to account for fermion masses and
mixings.

In the supersymmetric model, Dirac masses are generated by two distinct
Higgs doublets, $H_d$ and $H_u$, with VEVs $v_d$ and $v_u$,
through the Yukawa terms
\ini
Y_u Q H_u u+Y_d Q H_d d,
\fin
\ini
Y_{\nu} L H_u \nu+Y_e L H_d e,
\fin
where $Q$ and $L$ are quark and lepton weak doublets, and $u$, $d$, $\nu$, $e$
are weak singlets. We see that $M_u$ and $M_{\nu}$ are generated by the
same Higgs doublet $H_u$, that is $M_{u,\nu}=Y_{u,\nu}v_u$.
In a similar way $M_d$ and $M_e$ are generated by the Higgs
doublet $H_d$, that is $M_{d,e}=Y_{d,e}v_d$.
This fact could already suggest the assumption of an approximate quark-lepton
symmetry. The hierarchy (5) is obtained for $v_u \gg v_d$, keeping
all Yukawa couplings for the third generation of order 1, so that
$m_t \simeq v_u$ and $m_b \simeq v_d$.
We find that $Y_{u,\nu}$ and $Y_{d,e}$ are produced through two independent
Yukawa potentials, with breaking parameters $\epsilon_u$,
$\epsilon'_u$ and $\epsilon_d$, $\epsilon'_d$, respectively.
A Dirac potential is in the form \cite{bar2}
\ini
L \simeq v \left(\psi_3 \psi_3+\psi_3 \frac{\varphi^a}{M} \psi_a+
\psi_a \frac{S^{ab}}{M} \psi_b+\psi_a \frac{A^{ab}}{M} \psi_b \right),
\fin
where $\varphi^a$, $S^{ab}$, $A^{ab}$ are doublet, triplet and singlet flavon
fields, with VEVs given by
\ini
\frac{\varphi}{M} \simeq \left( \begin{array}{c}
\epsilon \epsilon' \\ \epsilon
\end{array} \right),~
\frac{S}{M} \simeq \left( \begin{array}{cc}
\epsilon'^2 & \epsilon \epsilon' \\ \epsilon \epsilon' & \epsilon
\end{array} \right),~
\frac{A}{M} \simeq \left( \begin{array}{cc}
0 & \epsilon' \\ -\epsilon' & 0
\end{array} \right),
\fin
and the mass $M$ is the cutoff scale of the effective theory, corresponding
to the existence of very massive states. The flavons are weak singlets.

The heavy neutrino mass is generated by a bare mass term
$M_R \nu \nu$
and is related to the $U(2)$ breaking involved in the up sector,
as expected since the $\nu$ fields appear. It is noteworthy that
we find this result from the independent input of large neutrino mixing.
Contrary to Dirac masses, which are generated at the weak scale,
the Majorana mass of the right-handed neutrino is not constrained and
can be very large, allowing the seesaw mechanism.
The Majorana potential looks like
\ini
L_R \simeq M_3 \left( \nu_3 \nu_3+ \nu_3 \frac{\phi^a}{M} \nu_a+
\nu_a \frac{\phi^a \phi^b}{M^2} \nu_b \right),
\fin
with the VEV of $\phi^a$ given by
\ini
\frac{\phi}{M} \simeq \left( \begin{array}{c}
\epsilon' \\ \epsilon
\end{array} \right),
\fin
and without the triplet (and of course the antisymmetric singlet) contribution.
We can also assume, in analogy to Eqns.(23),(24), that $M_R$ is generated
through a Yukawa term $Y_R \nu H \nu$ with a singlet Higgs $H$
having a very high VEV $v_R \simeq M_3$, in such a way to couple different
Higgs fields to different flavon fields.

The present model ascribes the hierarchy (5), and its analogue in the lepton
sector, to the coupling with two distinct
Higgs doublets, which is necessary in the supersymmetric case \cite{2h}, and the
hierarchies (1),(2), and their analogues in the lepton sector,
to the coupling with two corresponding
different sets of flavon fields, responsible for the $U(2)$ breaking.
As a consequence, the quark-lepton symmetry itself, in the form of
Eqns.(15),(16), could come out of this framework. Thus, an approximate
quark-lepton symmetry can be realized without the $SO(10)$ grand unification
or the $SU(4) \ts SU(2) \ts SU(2)$ partial unification, avoiding a possible
conflict with proton decay \cite{prode}. It is worth noting that the model
is in some sense economical with respect to unified theories,
since many gauge and Higgs bosons need not to be introduced,
while flavon fields have in any case to appear, in order to account for
the hierarchies. However, we do not exclude a possible realization
within the $SO(10)$ model. 

In summary, from quark-lepton symmetry, seesaw mechanism and the large lepton
mixing, we have inferred a simple model for fermion mass matrices in the
supersymmetric standard model with right-handed neutrinos and horizontal
$U(2)$ symmetry. Such a model suggests an alternative to the usual origin of
quark-lepton symmetry. The basic features of masses and mixings are produced
by VEVs of scalar fields. In particular, Higgs fields determine the overall
scales of mass matrices, and flavon fields their internal hierarchies.
The ratio $v_u/v_d = \tan \beta$ is predicted to be very large.

Finally, we discuss the implications for baryogenesis via leptogenesis
\cite{fy}. This mechanism for the generation of a baryon asymmetry $Y_B$
in the universe is based on the decay of the heavy right-handed neutrinos,
producing a lepton asymmetry which is partially converted into a baryon
asymmetry by sphaleron processes.
See Ref.\cite{ft} for a collection of the relevant formulas. Using the
approximate expression for the supersymmetric case and Eqns.(15),(17),
\ini
Y_B \sim \frac{M_1}{M_P} \frac{[(M_{\nu}^{\dg}M_{\nu})_{12}]^2}
{[(M_{\nu}^{\dg}M_{\nu})_{11}]^2} \frac{M_1}{M_2},
\fin
where $M_P \sim 10^{19}$ GeV is the Planck mass, we obtain
\ini
Y_B \sim \lambda^{12}~ \frac{M_3}{M_P}.
\fin
The required value $Y_B \sim \lambda^{15}$ is achieved for $M_3 \sim 10^{17}$
GeV, by one order higher than the unification scale, and corresponding to
$m_1 \sim 10^{-5}$ eV. This result has to be taken with care, due to the
approximation and the dependence on Yukawa coefficients.

\end{document}